\documentclass{elsart}
\usepackage{graphicx,harvard,amssymb}
\journal{New Astronomy}

\makeatletter
\def\elsartstyle{%
	\def\normalsize{\@setfontsize\normalsize\@xiipt{14.5}}
	\def\small{\@setfontsize\small\@xipt{13.6}}
	\let\footnotesize=\small
	\def\large{\@setfontsize\large\@xivpt{18}}
	\def\Large{\@setfontsize\Large\@xviipt{22}}
	\skip\@mpfootins = 18\p@ \@plus 2\p@
	\normalsize
}
\makeatother

\def\url#1{{\ttfamily\def\/{/\discretionary{}{}{}}#1}}

\def\mincir{\raise -2.truept\hbox{\rlap{\hbox{$\sim$}}\raise5.truept
\hbox{$<$}\ }}
\def\magcir{\raise -4.truept\hbox{\rlap{\hbox{$\sim$}}\raise5.truept
\hbox{$>$}\ }}

\begin{document}

\begin{frontmatter}

\title{Nature of Dark Energy and Polarization Measurements}

\author{R. Mainini\thanksref{rm}},
\author{L.P.L. Colombo\thanksref{lc}} \&
\author{S.A. Bonometto\thanksref{sb}}
\address{Physics Department G. Occhialini, Universit\`a degli Studi di
Milano--Bicocca, Piazza della Scienza 3, I20126 Milano (Italy)}
\address{I.N.F.N., Via Celoria 16, I20133 Milano (Italy)}

\thanks[rm]{E-mail: roberto.mainini@mib.infn.it}
\thanks[lc]{E-mail: loris.colombo@mib.infn.it}
\thanks[sb]{E-mail: silvio.bonometto@mib.infn.it}

\begin{abstract}
High sensitivity polarization measures, on wide angular scales, together 
with data on anisotropy, can be used to fix DE parameters. In this paper, 
first of all, we aim to determine the sensitivity needed to provide
significant limits. Our analysis puts in evidence that there
is a class of DE models that polarization measures can possibly exclude 
soon. This class includes models with DE due to 
a Ratra--Peebles (RP) potential. Using a likelihood 
analysis, we show that it is possible to distinguish RP models from 
$\Lambda$CDM and other dynamical DE models, already with the sensitivity 
of experiments like SPOrt or WMAP, thanks to their negative $TE$ correlation
at low--$l$, when the optical depth $\tau$
is sufficiently large. On the contrary, fixing the energy scale $\Lambda$ 
for RP potentials or distinguishing between $\Lambda$CDM and other 
DE potentials requires a much lower pixel noise, that no planned 
polarization experiment will achieve. While reviewing this paper after 
the referee report, the first--year WMAP data were released. WMAP finds 
large positive anisotropy--polarization correlations at low $l$; this 
apparently excludes DE models with RP potentials.

\end{abstract}

\end{frontmatter}

\section{Introduction}
The nature of Dark Energy (DE) is one of the main puzzles of modern cosmology.
DE was first required by SNIa data (see, e.g., Perlmutter et al. 1998, 
Riess et al. 1998), that indicate an accelerated cosmic expansion.
A flat Universe with $\Omega_m \simeq 0.3$ and $\Omega_b h^2 \simeq 0.02$ 
is also favored by an analysis of CMB and LSS observations (see, e.g., 
Tegmark et al., 2001; Netterfield et al, 2002; Percival et al., 2002; 
Efstathiou et al, 2002; Schuecker et al., 2003; Pogosian et al.,2003; 
here $\Omega_{m,b}$: matter, baryon density parameters; $h$: Hubble 
parameter in units of 100 km/s/Mpc; CMB: cosmic microwave background; 
LSS: large scale structure). 

The cosmic component allowing flatness,
in spite of $\Omega_m \simeq 0.3$, should not be observable in the 
number--of--particle representation, in the present epoch. 
If this component is a false vacuum, whose pressure and energy density 
($p_{DE}$ and $\rho_{DE}$) have ratio $w=-1$, a severe fine tuning 
of $\rho_{DE}$, at the end of the electroweak transition, is required;
in fact, before it, the vacuum energy density could be more than $10^{50}$ 
times greater than now. 
A possible alternative, apparently reducing fine tuning requirements, is 
that DE arises from a self--interacting scalar field $\phi$ (dynamical Dark 
Energy or {\it quintessence}; see, e.g., Ratra--Peebles 1988, RP hereafter; 
Wetterich 1988, 1995; Ferreira \& Joyce 1998; Perrotta \& Baccigalupi 
1999; Steinhardt et al. 1999; Zlatev et al. 1999; Albrecht \& Skordis 
2000; Amendola 2000; Brax, Martin \& Riazuelo 2000). For dynamical DE,
$$
\rho_{DE} = {{\dot \phi}^2\over 2 a^2} + V(\phi)~~,~~~~
p_{DE} = {{\dot \phi}^2\over 2 a^2} - V(\phi),
\eqno (1.1)
$$

$V(\phi)$ being the self--interaction potential, $a$ is the scale
factor and dots indicate derivatives in respect to the conformal time
$t$. The ratio $w$ becomes negative as 
soon as $\rho_k = {\dot \phi}^2/2 a^2 < V$. For $\rho_k/V \simeq 1/2$, 
it is $w \simeq -1/3$ and dynamical DE approaches open CDM. Still 
smaller $\rho_k/V$ ratios allow to approach $w= -1$ and a $\Lambda$CDM  
behavior. In order to work out $w$ and its time variations, Friedman 
equations, in association with the equation of motion of $\phi$, are to 
be integrated.

Solutions clearly depend on the shape of $V$, which, in principle,
is largely arbitrary. However, among potentials admitting a {\it tracker} 
solution, a particular relevance is kept by RP and SUGRA (Brax \& Martin 
1999, 2000) expressions
$$
V(\phi) = \Lambda^{4+\alpha}/\phi^\alpha
~,~~~~~~~~~
V(\phi) = (\Lambda^{4+\alpha}/\phi^\alpha) \exp (4\pi \phi^2/m_p^2)~,
\eqno (1.2)
$$
which originate within the frame of Supersymmetric (SUSY) theories. 
In the expressions (1.2), $m_p$ is the Planck mass and $\Lambda$ is an 
energy scale, currently set in the interval $10^2$--$10^{10}\, $GeV, 
ranging between the electroweak scale and the SUSY soft breaking scale. 
The potentials (1.2) also depend on the exponent $\alpha$; once $\Lambda$ 
and $\Omega_{DE}$ are assigned, however, $\alpha$ is univocally determined.
The RP and SUGRA potentials are also indicative of the opposite cases,
when $w$ shows slow or fast time variations, respectively.

Dynamical DE and $\Lambda$CDM predict quite similar observational
results, in most cases. This is a point in favor of dynamical DE,
as it is known that $\Lambda$CDM allows a good fit of most available
data. It is however important to devise suitable experiments whose
outputs enable us to probe DE nature; if DE is dynamical, they should
also enable us to fix the expression of $V$ and the values of the 
parameters on which $V$ depends.

Here we show that measures of anisotropy and polarization of CMB 
at large angular scales can constrain $V$. If favorable (but not 
exceptional) conditions are fulfilled, polarization measures, at a 
sensitivity level comparable with experiments in progress, can 
provide significant information, possibly {\it excluding} that a RP 
potential describes DE. 

We shall show this point by performing a likelihood analysis, assuming 
that polarization measures come from an experiment similar to the Sky 
Polarization Observatory (SPOrt: see, e.g., Macculi et al, 2000; Carretti 
et al, 2000; Peverini et al, 2001), considering both a noise level that the
experiment will attain and a lower noise level, as well. Our results, 
however, are not directly related to the above experiment and can be 
easily extrapolated to other observational contexts.

\section{The cosmic opacity to CMB photons}
In the present cosmic epoch, although diffuse baryonic materials are 
almost completely ionized, the scattering time for CMB photons $t_s 
\simeq 4.45 \cdot 10^{18} \Omega_b^{-1} h^{-2}$s exceeds the Hubble 
time $t_H = 3.09 \cdot 10^{17} h^{-1}$s and the Universe is transparent 
to CMB photons. These figures can be extrapolated to the past, but, while 
$t_s \propto a^3$, the dependence of $t_H$ on $a$ 
depends on the cosmological model. In an expansion regime dominated by 
non--relativistic matter, a fully ionized Universe is opaque to CMB at
redshifts  
$z > z_{op} \simeq 5\, (\Omega_b h)^{-2/3}$. Therefore, assuming 
$\Omega_b h^2 \simeq 0.022$ (standard BBNS limits; see, e.g., Dolgov 2002 
for a recent review) and $h \simeq 0.5$--0.7, we have $z_{op} > 64$, 
safely above any expected value for the reionization redshift $z_{ri}$. 
No reasonable cosmological model allows a substantial reduction of 
such $z_{op}$.

While, therefore, $\tau \ll 1$, its specific value depends on the 
model and, in particular, on the reionization history. If reionization 
is a single and sudden event, it can be synthesized by a value of $z_{ri}$. 
In Fig.~1 we show how $\tau$ is related to $z_{ri}$, for a wide set of 
cosmological models. In general, $z_{ri} \sim 8$ or $\sim 16$, if we 
require $\tau = 0.08 $ or 0.16. 

Data can constrain the reionization history in indirect ways.
High--$z$ QSO spectra (Djorgovski et al. 2001; Becker et al. 2001) 
show that at least a fraction of neutral hydrogen is present, in 
the intergalactic medium, at $z \gtrsim 6$. These very authors, however, 
warn the reader against an immediate conclusion that $z_{ri} \simeq 6$;
an ionization rate $x \sim 0.01$ can cause the Gunn--Peterson
effect observed. Their finding, however, was taken as an indication
of the expected range of $z_{ri}$.

Another constraint can be found, if the ionizing flux observed 
today is extrapolated to high $z$; in this way, Miralda--Escud\`e 
(2002) found that that the photon flux is able to reionize the cosmic
hydrogen at $z_{ri} \sim 8$--10 and $\tau \simeq 0.08$. 

Pushing farther the theoretical analysis, Tegmark (1997) estimated 
that, at $z \simeq 30$, a fraction $\sim 10^{-3}$ of baryonic matter 
is in fluctuations of $\sim 10^2$--$10^3 M_\odot$, already in non--linear 
growth. They can cool and turn into massive stars thanks to molecular 
hydrogen radiation. However, as soon as thermonuclear processes ignite,
stars produce photons hard enough to turn all hydrogen into atomic 
(or plasma) state and star formation is stopped until much greater
fluctuations ($\sim 10^7 M_\odot$) enter a non--linear regime
(see, e.g., Haiman 1997,  Gnedin \& Ostriker 1997). However, 
Haiman et al. (2000), Oh (2001) and Venkatesan et al. (2001)
find that the situation can be even more complicated, as X--ray 
heating might enhance the formation of molecular hydrogen.
From such pictures, Cen (2002) and Whyithe \& Loeb (2002) try
to deduce the reionization history. According to them, when the first 
PopIII stars form, a first high--$z$ reionization occurs. PopIII stars 
eventually produce metals that, in turn, allow less massive stars to 
form and, therefore, the ionizing photon flux is reduced soon. 
Then the overall ionization rate is set by a balance between star 
formation and hydrogen recombination; the former overwhelming the 
latter at a redshift $\sim 8$, when $x$ safely approaches unity.
At such redshift, this picture meets  Miralda--Escud\`e (2002) conclusions
and is therefore also coherent with the ionizing flux observed today.

These pictures are qualitatively realistic, although 
their quantitative aspects are still uncertain. It should be 
emphasized that different reionization histories, also when 
they yield the same $\tau$, give different CMB angular spectra.
In this work we shall quantitatively explore simple reionization 
histories, where a single reionization event yields $\tau = 0.08 $ or 0.16
and comment on the extension of our results to more complex histories.
The former $\tau$ value agrees with Miralda--Escud\`e (2002) estimate. 
The latter value assumes an earlier reionization, implicitly allowing 
a small hydrogen fraction ($\sim 1\, \%$) to remain neutral until 
$z \sim 6$, to account for the observed Gunn--Peterson effect in high--$z$
QSO spectra.

It should also be reminded that, until CMB spectra analysis is restricted to
anisotropy data, no significant constraint on $\tau$ can be set; Stompor
et al (2002), using balloon data, can only set the limit $\tau < 0.40$, at 
the 2--$\sigma$ level. Limits are so loose because of the degeneracy 
between $\tau$ and $n$ (primeval spectral index), in CMB anisotropy data. 
In a recent paper, Colombo \& Bonometto (2003) discussed how much such 
degeneracy can be removed using polarization measures. Recent 
observations (Kovac et al. 2002) apparently confirmed that CMB is 
polarized, but the sensitivity level and the angular scales probed by
such observations were not
adequate to reduce the above $n$--$\tau$ degeneracy.

In this work we explore what large angular scale
CMB polarization data can tell us on DE nature. 
For the sake of simplicity, we assume 
that all model parameters are obtained from measures at smaller 
angular scales and concentrate on the effect of the potential 
$V(\phi)$ on large angular scales, when $\tau$ is large.

\section{CMB angular spectra}
The linear evolution of density fluctuations in models with
a DE component yields the angular spectra $C_l^{A}$; 
$A=T,~E,~B,~X$ stand for anisotropy, $E$-- and $B$--mode 
polarization and anisotropy--polarization cross correlation (X=TE)
only $E$--mode polarization is considered through this paper. 
In Fig. \ref{2fig}--\ref{5fig} we show the 
spectra $C_l^{T,X,E}$ when the potential
is RP or SUGRA and $\tau$ is 0.08 or 0.16, for a set of values of the energy
scale $\Lambda$. All spectra are obtained with suitable 
extensions of CMBFAST
\footnote{\url{http://physics.nyu.edu/matiasz/CMBFAST/cmbfast.html}} 
. 

Figs. \ref{2fig} to \ref{5fig} show that the 
nature of DE and, in particular, the value
of the energy scale $\Lambda$, both for RP and SUGRA models, display 
their main effects on low--$l$ TE correlation spectra. 
In particular, Figs.\ref{2fig} and \ref{3fig}
show that $C^{X}_l$ is negative, 
at low $l$, for RP potentials, while, in SUGRA models, such feature
is absent. Negative $C^{X}_l$ are not an exceptional feature: all 
models yield alternate intervals of positive and negative 
$C^{X}_l$ as $l$ increases. In most spatially flat models, however, the 
$C^{X}_l$ spectrum, at low $l$ values, starts from positive values. On 
the contrary, open CDM models, when $\tau$ is large enough, start
with negative $C^{X}_l$, al low $l$.

In the discussion section, we shall discuss why RP models share this feature 
with open models. This arises from the way how $\rho_{DE}$ scales with $a$,
similar to the scaling of the curvature term in open models.
On the contrary, in $\Lambda$CDM models, $\rho_{DE}$ does not depend on 
$a$. When the scaling of $\rho_{DE}$ with $a$ is mild and, therefore,
close to $\Lambda$CDM, no low--$l$ negative $C^{X}_l$ is expected.

\section{Likelihood analysis}
\subsection{The SPOrt experiment}
In this section we aim to determine the likelihood of different models 
in respect to possible results of anisotropy and polarization measurements, 
obtained through definite observational apparati, for specific cosmological 
models. In particular, polarization data sets are built with reference 
to the features of the SPOrt experiment. 

At variance from other space experiments (WMAP and PLANCK), SPOrt has been 
explicitly designed to measure the Stokes parameters Q and U and to 
minimize systematics and instrumental polarization. SPOrt will observe 
a sky area with declination $|\delta | \le 51.6^o$ ($\sim 80\%$ of the 
whole sky). The lifetime of the experiment will be at least 1.5 years, but 
an extension is likely.

Here we assume that pixels are distributed according 
to the HEALPix\footnote{\url{http://www.eso.org/science/healpix/}} 
package, with $N^{side} = 16$, and smoothed with a Gaussian beam of FWHM 
of $7^o$. We therefore have, on the whole sky, 3072 pixels, whose centers 
lie at an average angular distance of $\sim 3.7^o$. Once polar caps are 
excluded, there remain 2448 pixels, providing measures of the Stokes 
parameters $Q$ and $U$.

SPOrt polarimeters will provide no anisotropy data. Such information 
will be however available through other experiments (COBE, WMAP).
Of course, anisotropy data probe a different sky area, i.e. the whole 
sky, excluding the area with galactic declination $| \delta | < 20^o$, where 
contamination by the Milky Way is severe. (On the contrary, such contamination 
can be assumed to be under control, for polarization data; see, e.g., 
Bruscoli et al. 2002; Tucci et al. 2002.) In this sky area 
there are 1984 pixels. For 1360 of them, both anisotropy 
and polarization data are supposed to be available.

Random noise ($\sigma^P_{pix}$) is included in artificial data, 
assuming it to be uncorrelated, both among pixels and between $Q$ and $U$, 
in the case of polarization measures. We assume $\sigma^T_{pix} = 1\, \mu$K 
for temperature data (as expected for WMAP measurements, extended to 4 
years, scaled to the SPOrt resolution). On the contrary,
we consider different noise levels for polarization data.

At present, the SPOrt experiment team expects
a pixel noise level of 

\vskip0.2truecm
\centerline{
$\sigma^P_{pix} \simeq 1.8\,{N^{side} \over 8}\, 
\sqrt{{2 \over \lambda} {0.6 \over \epsilon}}$ $\mu$K, 
}

where $\lambda$ is the experiment's lifetime in years while $\epsilon$ 
indicates the detection efficiency. Our analysis considers 
$\sigma^P_{pix}$'s in the interval comprised from 0.1 to 1$\, \mu$K per 
pixel. Even at its highest value, this noise is consistent with SPOrt
claims only if $\lambda = 4\, y$ and efficiency approaches unity. 
However, the SPOrt payload is to be delivered in more than one year 
and there is room for some further improvement in the apparati.

\subsection{Models explored}
In this analysis we take $\Omega_{DE} = 0.7$, $\Omega_b h^2 = 0.022$, 
$h=0.7$ and a primeval spectral index $n=1$. We considered also 
different values for these parameters, but the likelihood distributions 
are essentially independent from them. On the contrary, such distributions 
depend on the choice of $\tau$; as already outlined, we studied likelihood 
distributions for $\tau=0.08$ and 0.16$\, $. 

We consider first RP models, for $\Lambda/$GeV$= 10^3$ and $10^6$. 
As is known, RP models are accelerated only if $\Lambda \mincir 10^4$GeV
and disagree with SNIa data at the 2--$\sigma$ level, when $\Lambda \magcir
10^6$GeV. On the contrary, when $\Lambda \mincir 1\, $GeV, 
the angular spectra of the model are hardly distinguishable from $\Lambda$CDM.
Accordingly, fits yielding $\Lambda \mincir 1\, $GeV, for a RP model, indicate 
that data do not allow to appreciate that the model is different from 
$\Lambda$CDM.

We consider then SUGRA models, for $\Lambda/$GeV$= 10^3$, $10^6$
and  $10^9$. Their $TE$ correlation spectra show differences from 
$\Lambda$CDM and among themselves, although not so relevant as for RP.

$TE$ correlation
spectra also depend on the reionization history. In this paper,
such dependence is not explored and this is the main quantitative
limitation to the results shown here.

Dealing with low $l$'s, cosmic variance must be carefully taken into 
account. We explore this point by simulating and analyzing a large number 
of artificial data sets. Results will be both synthesized into average 
predictions and analyzed providing the frequency of possible determinations.

\subsection{Results}
We report first results for RP models, for $\Lambda/$GeV$= 10^3$ and $10^6$. 
In Figs.~6 and 7 we show the average likelihood distributions for such models.
In Figs. \ref{9fig} and \ref{10fig}, we report the distributions of 
the 1--2 $\sigma$
lower limits on $\Lambda$, for $\tau=0.08$ and 0.16, $\Lambda=10^3$ and
$10^6$, $\sigma^P_{pix} = 0.1\, \mu$K and 1$\, \mu$K.


Fig. \ref{12fig} is analogous to Fig. \ref{6fig}, but concerns a 
SUGRA model with 
$\Lambda=10^9$GeV and $\tau = 0.16$; among the cases considered here, 
this is the one which allows the best recovery of $\Lambda$.
In Fig. \ref{13fig}, we report the distributions of the lower limits 
on $\Lambda$ for such model (at 1$\, \sigma$).

The above figures show that the energy scale $\Lambda$ cannot be
detected for SUGRA models, at least with large--angular--scale
experiments. This is not only true at a noise level achievable by 
SPOrt, but also when noise is reduced by a factor $\sim 10$.

The situation was better for RP models. The capacity of an
experiment to distinguish a RP model from $\Lambda$CDM
increases with the energy scale $\Lambda$ and the opacity 
$\tau$, although being more sensitive to the former parameter. 
In fact, a substantial fraction of realizations, with 
$\Lambda=10^6$GeV and $\tau=0.08$, can be distinguished from
$\Lambda$CDM, even at the 3--$\sigma$ level; such fraction is
only slightly greater for $\tau = 0.16$. Furthermore, only in quite 
a small fraction of realizations, RP cannot be distinguished from 
$\Lambda$CDM, at the 2--$\sigma$ level.
Both with $\tau = 0.08$ and 0.16, also models with $\Lambda = 10^3$GeV 
can be distinguished from $\Lambda$CDM in a smaller (but still substantial) 
fraction of realizations, at least at the 2--$\sigma$ level.


If we aim farther and try to determine $\Lambda$, we meet substantial
difficulties: even reducing the noise level by a factor $\sim 10$, 
the best we can find is a lower limit on the order of magnitude of
$\Lambda$. Unfortunately, this can be fixed just in a fraction of 
realizations.

\section{Discussion}
Let us first discuss how the likelihood distribution can be obtained 
from the spectra of a given model, taking into account that the number 
of pixels for anisotropy and polarization ($N_T$ and $N_P$) in our 
(artificial) data are however different. Let then $T_j$ be the anisotropy 
measured in $N_T$ pixels and $Q_j$ and $U_j$ the Stokes parameters measured 
in $N_P$ pixels. In general, let us define vectors ${\bf x} \equiv 
(T_1,.....,T_{N_T}, Q_1,.....,Q_{N_P}, U_1,.....,U_{N_P})$, of $N_s = 
N_T + 2 N_P$ components, defining an observed state of anisotropy and 
polarization. Once a $\Lambda$ value is assigned, 
the angular spectra $C_{l}^A= ( C_{l}^T,C_{l}^E,C_{l}^{X} )$ are 
univocally determined. On the contrary, a data vector ${\bf d}$, of 
$N_s$ components, built from them, is just a {\it realization} of such 
model: once the $N_s$ component vector ${\bf d}$ is assigned, the value 
of $\Lambda$ is not univocally fixed.

A function
$$
{L}({\bf d}|C^A_l) \propto [\, \det {\bf M}\, ]^{-{1 \over 2}} \exp 
\big[ -{1 \over 2}{\bf d^T} {\bf M}^{-1} {\bf d} \big]
\eqno (5.1)
$$
is then built, to yield the likelihood of a given set of $C^A_l$ 
(i.e., of a given $\Lambda$ value), if ${\bf d}$ is observed.
The main ingredient of $L$ is the correlation matrix 
${\bf M_{ij}} = \langle {\bf x}^T_i {\bf x}_j \rangle = {\bf S}_{ij} 
+ {\bf N}_{ij}$; here ${\bf S}_{ij} $ is the signal term and
${\bf N}_{ij}$ is due to the noise. The components ${\bf M_{ij}}$
yield the correlation between the $i$th and $j$th elements of 
data vectors ${\bf x}$ corresponding to particular choices of $C_{l}^A$,
i.e. of $\Lambda$ values. The construction of the (model dependent) 
signal term, however, does not require to build explicitly the vectors 
${\bf x}$. The procedure to be followed, in the case when both 
anisotropy and polarization data are available, is explicitly
reported by Zaldarriaga (1998). The construction of the noise term 
is simpler, as we expect no noise correlation, and the matrix 
${\bf N}_{ij} = \delta_{ij}\,{\sigma_{T,pix}}^2$ (for $i = 1,...,N_T$) 
and ${\bf N}_{ij} = \delta_{ij}\,{\sigma_{P,pix}}^2$ (for $i = 
N_T+1,...,N_s$) is diagonal.

In what follows, the technical role of the $C_l^A$ spectra does not
need to be further outlined and the likelihood function is explicitly 
considered to depend on $\Lambda$. In particular, we assume that the 
most probable $\Lambda$ value, for a given anisotropy--polarization 
state ${\bf d}$, is the one which maximizes the likelihood. By 
integrating the likelihood distribution along the $\Lambda$ axis, 
we are then able to find the intervals corresponding to 1--2--3$\, 
\sigma$ levels.

In order to explore cosmic variance, 1000 realizations of each model 
were considered; the likelihood curves plotted in the Figs. 5--6--7--11 are
averages of the results of such realizations. 1--2--3$\, \sigma$ value 
distributions, instead, allow to appreciate the spread among the
results of each such realization.

The feature which explains why RP model likelihoods can enable us to 
distinguish them from $\Lambda$CDM, in several cases, is the negative 
TE correlation at low $l$. In the case of SUGRA models such TE 
anticorrelation is absent and large angle data seem unsuitable to 
provide information on DE nature.

Negative TE correlations are essentially related to the simultaneous 
action of ISW effect and cosmic opacity. The former effect is a consequence
of the varying rate of expansion, when we pass from matter to curvature 
dominance, in open models, or from matter to vacuum dominance in 
$\Lambda$CDM models. However, while the former kind of passage, 
in the presence of cosmic opacity, does produce anticorrelation, 
the latter kind of passage does not. 

The capacity of RP models to induce anticorrelation, as we shall see,
is related to their features, which more closely approach open CDM,
rather than $\Lambda$CDM. The opposite is true for SUGRA models.

A full understanding of model dynamics requires that the whole set of 
Boltzmann equations, describing photon distribution, is followed in time. 
Such equations are shown in a number of papers, together with equations 
ruling the dynamics of fluctuations in other components of a model (see, 
e.g., Ma \& Bertschinger 1995). Signs and other definitions that we 
adopt here are the same as in the public code CMBFAST. Let us then indicate 
by $F_l(k,t)$ and $G_l(k,t)$ the Boltzmann components for anisotropy
and polarization, respectively, $k$ and $t$ being the wave--number and 
the conformal time. 

Let then $n_e$ and $\sigma_T$ be the free electron density and
the Thomson cross--section. The opacity to CMB photons 
reads then $\tau = \int_t^{t_o} n_e(t') \sigma_T a(t')\, dt'$ and
$-{\dot \tau}=a\, n_e \sigma_T $.

Let us report then the set of equations fulfilled by the $G_l$ components
in spatially flat models:
$$
\dot G_l = -{\dot \tau} \big[-G_l + {\Pi \over 2}\big( \delta_{l0}
+{\delta_{l2} \over 5}\big)\big] + {k \over 2l+1} [l\, G_{l-1}
- (l+1) G_{l+1} ] ~
\eqno (5.2f)
$$
($\delta_{ln}$ is the Kronecker symbol). Here
$$
\Pi = G_o + G_2 + F_2,
\eqno (5.3)
$$
is the only vehicle transferring signals from anisotropy to polarization. 

In a non-flat geometries eq. (5.2f) becomes
$$
\dot G_l = -{\dot \tau} \big[-G_l + {\Pi \over 2}\big( \delta_{l0}
+{\delta_{l2} \over 5}\big)\big] + {\beta \over 2l+1} [l b_l\, G_{l-1}
- (l+1) b_{l+1} G_{l+1} ] ~;
\eqno (5.2o)
$$
here $\beta^2 = k^2 + K$ and $b_l^2 = 1 - K l^2/\beta^2$, with $K =
-(1-\Omega_m) H_0^2$, $H_0$ being the present value of the Hubble
parameter.

Quite in general, when $n_e$ and, therefore, $-\dot \tau$ is large, 
all $G_l$ are exponentially damped. On the contrary, when $n_e$ is 
low, the polarization signal spreads along the $G_l$ component series. 

In order to produce polarization, therefore, there must however be
a seed coming from anisotropy, linked to the non--vanishing of
the quadrupole term $F_2(k,t)$. If $F_2(k,t)$ becomes significant only 
when $n_e$ is very low, little polarization is produced on harmonics with
$l$ corresponding to such $k$. It is so for wavelengths $2\pi/k$ 
entering the horizon well after recombination, which correspond to 
spectral components with $l \ll 200$, unless reionization occurs. 
This explains the sudden rise of $C^P_l$ in Fig.~2--5, when $\tau 
> 0$ is considered.

Let $F_l^o$ ($G_l^o$) be the value taken by $F_l$ ($G_l$) at the
present time $t_o$. In spatially flat models, the angular spectra
read
$$
C^T_l = {\pi \over 4} \int d^3k\, P_o(k) \, \left|{F_l^o(k) }\right|^2 ~, ~~
C^P_l = {\pi \over 4} \int d^3k\, P_o(k) \, |G_l^o(k)|^2 ~, $$$$
C^{X}_l = {\pi \over 4} \int d^3k\, P_o(k) \, {F_l^o(k)} G_l^o(k) ~,
\eqno (5.4f)
$$
$P_o(k)$ being the primeval fluctuation spectrum. In open model,
angular spectra expressions are slightly more complex and read
$$
C^T_l = {\pi \over 4} \int d^3\beta\, P_o(q) \, \left|{F_l^o(\beta)}
\right|^2 ~, ~~
C^P_l = {\pi \over 4} \int d^3\beta\, P_o(q) \, |G_l^o(\beta)|^2 ~, $$$$
C^{X}_l = {\pi \over 4} \int d^3\beta\, P_o(q) \, {F_l^o(\beta)} G_l^o(\beta) ~;
\eqno (5.4o)
$$
here $q= (\beta^2-4K)^2/\beta(\beta^2 -K)$. Clearly, for $\Omega_m=1$,
both $\beta$ and $q$ return $k$.

Both eqs. (5.2) and (5.4) indicate the presence of suitable shifts
in the $k$--space, when open models are considered. In particular,
the $b_l$ coefficients in eq.~(5.2o) are responsible for the shifts
of $C_l$ peaks, while the shifts due to the passage from $k$ to
$\beta$ and $q$, in eq.~(5.4o), displace the power of the spectrum
$P_o$ through the harmonics $F_l$ at small $l$, i.e. on scales 
comparable with the curvature scale. 

Apart of these geometric and power shifts, the main difference between 
flat and open models resides in the equation fulfilled by the gravitational
field fluctuations $\dot h$ and $\eta$. In the synchronous gauge, 
for open models:
$$
2{\bar k}^2 {\dot \eta} = 8\pi G a^2 [\sum_c (\rho_c+p_c) \theta_c 
- {\dot h} \rho_{o,cr} (1-\Omega_m)/a^2] ;
\eqno (5.5o)
$$
here $\rho_{o,cr}$ being the critical density at $z=0$ and $\sum_c$ 
indicates a sum over all (relativistic or non--relativistic) matter 
components. In the same gauge, for flat models with DE:
$$
2k^2 {\dot \eta} = 8\pi G a^2 [\sum_c (\rho_c+p_c) \theta_c + 
(\rho_{DE}+p_{DE}) \theta_{DE}] ,
\eqno (5.5f)
$$
where the DE term has been set in evidence, with its possible
fluctuations. The wave--numbers $\bar k^2$ and $k^2$ are shifted by 
$3 K$. Once again this corresponds to a shift in the $k$--space.

Let us notice soon that, in a $\Lambda$CDM model, no DE fluctuations 
exist and, furthermore, $\rho_{DE} = - p_{DE}$, so that the second 
term at the r.h.s. of eq.~(5.5f) vanishes.

In models with dynamical DE, instead, $\theta_{DE}$ does not vanish and, 
in general, $\rho_{DE} \neq - p_{DE}$. Accordingly, the second term in 
square brackets in eq.~(5.5f) may read $\theta_{DE} \rho_{o,cr} (1-
\Omega_m)(1+w) (\rho_{DE}/\rho_{o,DE})$ (the ratio in the last 
parenthesis tells us how DE energy scales with $a$). This term is then 
analogous to the second term in square bracket in eq.~(5.5o). It would 
coincide with it if $w=-1/3$ and, namely, if $\theta = -{\dot h}/2$.
If the latter is true, we can expect that, apart of a different power 
distribution along the $l$ axis and geometric effects at greater $l$,
there can be specific similarities in 
the behavior of open and dynamical DE models.

The relation between $\dot h/2$ and $\theta_{DE}$ can be studied
through the equation
$$
\theta_{DE} + {{\dot h} \over 2} = - {1 \over 1+w}\, \big[{\dot \delta_{DE}}
+ 3 {\dot a \over a} (c_s^2-w) \delta_{DE}\big],
\eqno (5.6)
$$
whose validity indicates that DE behaves as a fluid; here 
$c_s$ is the sound velocity in DE. This equation can be
derived from the equation of motion of $\phi$, as shown in Appendix A. 
An order of magnitude estimate, however, tells us soon that $\theta_{DE} 
\sim k^2 t\, \delta_{DE}$; then, the ratio between $\theta_{DE}$ and 
the r.h.s. is $\sim (t/L)^2$, where $L$ is the scale related to $k$. 
When $t \ll L$ (before horizon crossing), the $\theta_{DE}$ term is 
negligible, in comparison with the r.h.s.. For $L \ll t$ (after horizon 
crossing), the contrary is true. Hence, before horizon crossing, 
$\dot h/2$ is essentially required to equate the r.h.s. and, therefore, 
the ratio $-\dot h/2 \theta_{DE}$ shall exceed unity. At horizon 
crossing such ratio must approach unity and keep to such value as 
$t$ grows greater than $L$.

This point can also be numerically inspected. In Fig. \ref{14fig}, the ratio 
$-{\dot h}/2 \theta_{DE}$ is plotted against $a$, for those $k$ values 
which yield the top contribution to the $l$ harmonics indicated in the 
figure, in the case of a RP model. Clearly, about horizon crossing, 
$\theta_{DE}$ is already quite close to $-\dot h/2$. 

The main differences between open and dynamical DE models, in the r.h.s. 
of eqs.~(5.5) are therefore relegated to times before horizon crossing. 
Afterwards, the residual difference is due to a factor $1+w$. If 
$w$ is too close to $-1$, however, such factor risks to spoil the 
similarities noticed hereabove.

The changes in $\dot \eta$ directly act on $F_2$. The equation fulfilled 
by this spectral component reads:
$$
\dot F_2(k,t) = a\, n_e \sigma_T \left[-F_2(k,t) 
+ {\Pi(k,t) \over 10}\right]  
+ {k \over 5} [2\, F_1(k,t) - 3 F_3(k,t) ] + $$$$
+{8 \over 5} \left({\dot \eta(k,t)}
+ {{\dot h(k,t)} \over 6} \right)
\eqno (5.7f)
$$
in flat models, while in open models we have:
\vskip 0.2truecm
$$
\dot F_2(\beta,t) = a\, n_e \sigma_T \left[-F_2(\beta,t) 
+ {\Pi(\beta,t) \over 10}\right] 
+ {\beta \over 5} [2 b_2\, F_1(\beta,t) - 3 b_3 F_3(\beta,t) ] +$$$$ 
+{8 \over 5}  {\beta \over k} b_2
 \left({\dot \eta(\beta,t)}
+ {{\dot h(\beta,t)} \over 6} \right) ~.
\eqno (5.7o)
$$
Let us recall that $F_2$ is the only term of $\Pi$ where anisotropy
contributes. Such contribution is therefore directly affected by
a change in the equation of the gravitational field, which
describes the ISW effect in the synchronous gauge and can be
similar in open models and in models with dynamical DE, provided
that $w$ is not too far from -1/3.

Accordingly, when open models show negative TE correlations at low $l$,
we can expect the same to occur in flat RP models. On the contrary, 
SUGRA models, which grant a cosmic acceleration much closer to 
$\Lambda$CDM models, due to a $w$ value not too far from -1, should 
not be expected to give negative TE correlation and cannot be tested 
in the same way as RP models.

\section{Conclusions}
In this paper we showed that a particular class of dynamical DE models 
can be fairly safely {\it falsified} using large angle CMB polarization 
data. Models with DE due to a scalar field, self--interacting through 
a RP potential, belong to such class. In general, however, we expect
that any model with dynamical DE, yielding $w \sim -1/3$ or just
slightly smaller, can be tested through similar data.

Our treatment provides quantitative predictions for the case of
RP potentials, with the simplest possible reionization history
yielding $\tau = 0.08$ or 0.16. If the reionization history was
more complicated, as is however likely, quantitative conclusions
do not apply and is hard to parametrize the whole functional
parameter space, providing precise predictions. An analysis of
$TE$ correlation spectra, however, shows that negative values
are a generic feature, at low $l$, for most reionization histories.
Accordingly, in the absence of such negative signal, a RP model is 
however disfavored.

An experiment to detect CMB polarization with a resolution of $\sim 7^o$
and $\sigma^P_{pix} \simeq 1\, \mu$K is fully realistic. With such noise, 
there is a substantial fraction of realizations of a RP model
with $\Lambda=10^6\, $GeV, with $\tau=0.16$, which can be distinguished 
from $\Lambda$CDM at the 3--$\sigma$ level. Accordingly, the 
data analysis of a polarization experiment, with a noise level of
$\sim \sigma^P_{pix} = 1\, \mu$K, has, at least, $\sim 30\, \%$ of 
probability to indicate that the model is RP, if this is true. 
At the 2--$\sigma$ level, such model could be confused with 
$\Lambda$CDM only in quite a small fraction of realizations. 



The only planned polarization experiment specifically aiming to large 
angle polarization data is SPOrt. The noise level that SPOrt now expects
is not much more than 1$\, \mu$K per pixel. Other experiments, planned to 
provide data with greater resolution, can also provide large--angle
harmonics. For instance, the WMAP satellite, although not built to 
minimize systematic effects on polarization, can be able to provide 
polarization data with a noise level just slightly worse than SPOrt,
so allowing an earlier test of RP models. Finally, with the sensitivity 
expected for PLANCK, RP and similar models can be surely tested.

In order to determine the energy scale $\Lambda$ for RP potentials or 
to distinguish between $\Lambda$CDM from other DE potentials, using
large angular scale data, a pixel noise even below 0.1$\, \mu$K is 
apparently required. No planned polarization experiment will achieve 
such noise level soon and we can expect that an insight on the shape 
of DE potentials can come earlier from different kinds of experiments.

\section{Note added after WMAP release}
After submission of this paper and during its revision following the
referee report, the first release of WMAP data occurred
\footnote{\url{http://lambda.gsfc.nasa.gov/product/map/m\_overview.html}}.
This suggests to add some further comments. 

Apparently WMAP data give no indication of $C^{X}_l$ being negative.
The three procedures, used to estimate it, all yield large $C^{X}_l$ 
values at low $l$ and $\tau = 0.16 \pm 0.04$, for $\Lambda$CDM models
and a single reionization. One of the above procedures explicitly 
assumes that a fair approximation to the true model can be found among
$\Lambda$CDM models, trying to fix $\Lambda$CDM parameters best fitting data. 
The $C^{X}_l$ spectrum was also related its Fourier transform, and then 
determined with a third procedure claimed to be fully model--independent.

Here we wish to outline that the first procedure is unsuited to 
exclude RP models. Following the pattern of Sec.~5, we created a set of 
artificial skies for RP models with $\Lambda = 10^3$GeV and $10^6$GeV and 
$\tau = 0.16$. We applied the maximum likelihood procedure to 
these data and fit cosmological parameters, {\it assuming that 
the true model is} $\Lambda$CDM. For the sake of example, in Fig. 
\ref{ll3fig} 
we show the likelihood distribution on $\Omega_\Lambda$, 
for a RP model with $\Lambda = 10^3$GeV.
The procedure, being unable to explore a fair class of models,
systematically overestimates $\Omega_{DE}$; if the noise
is $\sim 1\, \mu$K, the likelihood distribution is significantly
non-Gaussian. When noise is reduced, the likelihood distribution 
is more Gaussian, but still peaks at high $\Omega_{DE}$.
Then, for $\sigma^P_{pix} = 0.1 \,\mu$K, the Gaussian
becomes very narrow and the fair $\Omega_{DE}$ value lies at 
more than 3--$\sigma$ from the best fitting $\Omega_{DE}$.
The situation is similar for $\Lambda = 10^6$GeV.

In Fig. \ref{cxlfig}, we make use of this latter model
to compare the RP spectrum used to create data, with the 
best--fit $\Lambda$CDM spectrum. The continuous line 
connects the $C_l^{X}$ values of the RP model; the long--dashed 
line yields the $C_l^{X}$ values for a $\Lambda$CDM model with 
the right $\Omega_{DE}$; the short--dashed connects the 
$C_l^{X}$ values for the best--fit $\Lambda$CDM model, with 
1--$\sigma$ error--bars. The shaded area shows the range of the
cosmic variance. 

For the sake of comparison, Figs. \ref{llsufig} and \ref{cxlsufig} 
are analogous to Figs. 
\ref{ll3fig} and \ref{cxlfig}, but for a SUGRA
model. In the absence of the peculiar negative behavior, characterizing
the RP case, no misleading effect is present.

With the reserves that these tests suggest, and relying
on the {\it model independent} $C_l^{X}$ deduction procedure,
we however conclude that WMAP apparently excludes
that DE is due to a scalar field, self--interacting through a RP 
potential with $\Lambda \magcir 1$--10$\, $GeV. We plan to deepen
the quantitative aspects of this finding in a forthcoming paper.

Acknowledgements -- LPLC acknowledges the financial support of ASI,
within the activities related to the SPOrt experiment. The CINECA
consortium is thanked for making its facilities available for
this work. The components of the SPOrt team are also thanked for
discussions on the topics of this research. The public programs 
CMBFAST, by U. Seljak \&
M. Zaldarriaga, and HEALPix, by K.M. G\`orski et al. were widely 
used in the preparation of this work.

\newpage

\section*{Appendix A}
Starting from the Lagrangian density for the $\phi$ field,
self--interacting through a potential $V(\phi)$, and using 
Eulero--Lagrange equations, we can work out the equations for the
background unperturbed field $\phi$ and its fluctuations $\delta \phi$. 
The latter reads
$$
\ddot {\delta \phi} + 2 {\dot a \over a} \dot {\delta \phi}
+ (k^2 + V''a^2)\, \delta \phi + \dot \phi {\dot h \over 2} = 0
\eqno (A1)
$$
and we aim to show that this is equivalent to eq.~(5.6), that
we report here:
$$
\theta + {\dot h \over 2} = -{1 \over 1+w} \big[ \dot \delta
+ 3 {\dot a \over a} (c_s^2 - w) \delta \big] ~.
\eqno (A2)
$$
An equation equivalent to eq.~(A2) is given, for a generic
fluid, by Ma \& Bertschinger (1955), who derive it from the
pseudo--conservation of the energy--momentum tensor $T_{ij}$. 

In this Appendix, for the sake of simplicity, we omit the subscript 
$_{DE}$ to 
$$
\delta = {\delta \rho_{DE} \over \rho_{DE}} ~~~~ {\rm and}
~~~~~~
\theta = {k^2 \over a^2} {\dot \phi\, \delta \phi \over
\rho_{DE}(1+w)} ,
\eqno (A3)
$$
as well as to the energy density $\rho$ and to the pressure $p$
of DE.

The proof is simpler if we start from eq.~(A2) and obtain eq.~(A1).
Let us first multiply both sides of eq.~(A2) by $(1+w)\rho$ and replace
$\dot \delta$ by 
$$
{\dot {\delta \rho} \over \rho} - {\dot \rho \over \rho} {\delta \rho
\over \rho} = 
{\dot {\delta \rho} \over \rho} + 3{\dot a \over a} (1+w) {\delta \rho
\over \rho} ~.
\eqno (A4)
$$
In this way, we obtain
$$
\rho (1+w) \big(\theta + {\dot h \over 2} \big)
= -\dot {\delta \rho} - 3 {\dot a \over a} (1+c_s^2) \delta \rho ~.
\eqno (A5)
$$
Owing to the definition of $\theta$ (eq.~A3) and thanks to the relation
$(1+w)\rho = \rho+p=\dot \phi^2/a^2$, the l.h.s. of eq.~(A5) becomes
$(k^2/ a^2) \dot \phi\, \delta \phi + (\dot \phi^2 / a^2)
(\dot h/ 2) $, while, at the r.h.s., we may replace
$\delta \rho $ and $\dot {\delta \rho}$ with the expressions
obtainable from the definition of $ \rho = \dot \phi^2/2a^2 + V$.
Both sides of the resulting equation must then be divided by
$\dot \phi/a^2$, so obtaining
$$
k^2 \delta \phi + \dot \phi {\dot h \over 2} =
-\ddot {\delta \phi} - \big({\ddot \phi \over \dot \phi}
+ a^2 {V' \over \dot \phi} - 2 {\dot a \over a}
\big) \dot {\delta \phi} - V'' a^2 \delta \phi
-3 {\dot a \over a} (1+c_s^2) \big( {\dot \phi \dot {\delta \phi}
\over a^2} + V' \delta \phi \big) {a^2 \over \dot \phi} ~.
\eqno (A6)
$$
Here we ought to use the equation of motion of the unperturbed 
field $\phi$, which can be easily reset to yield that 
$\ddot \phi/\dot \phi = -2\dot a/a - a^2V'/\dot \phi$. 
Rearranging the terms, we then obtain:
$$
\ddot {\delta \phi} + 2 {\dot a \over a} \dot {\delta \phi}
+ (k^2 + V''a^2)\, \delta \phi + \dot \phi {\dot h \over 2} = 
3{\dot a \over a} \big[(1-c_s^2) \dot {\delta \phi} -
(1+c_s^2) a^2 {V' \over \dot \phi} \delta \phi \big] ~.
\eqno (A7)
$$
Eq.~(A7) coincides with the equation of motion (A1), provided that
the r.h.s. vanishes. This follows the very definition of
sound velocity
$$
c_s^2 = {\delta p \over \delta \rho}
= {\dot \phi \dot{\delta \phi}/a^2 - V' \delta \phi
\over \dot \phi \dot{\delta \phi}/a^2 + V' \delta \phi}
\eqno (A8)
$$
from which we easily obtain that $(1+c_s^2)/(1-c_s^2)=
\dot \phi \dot {\delta \phi}/a^2 V' \delta \phi$.
This last relation clearly yields the vanishing of the r.h.s. of eq.
(A7).

\newpage

\begin{figure}
\begin{center}
\includegraphics*[width=9cm]{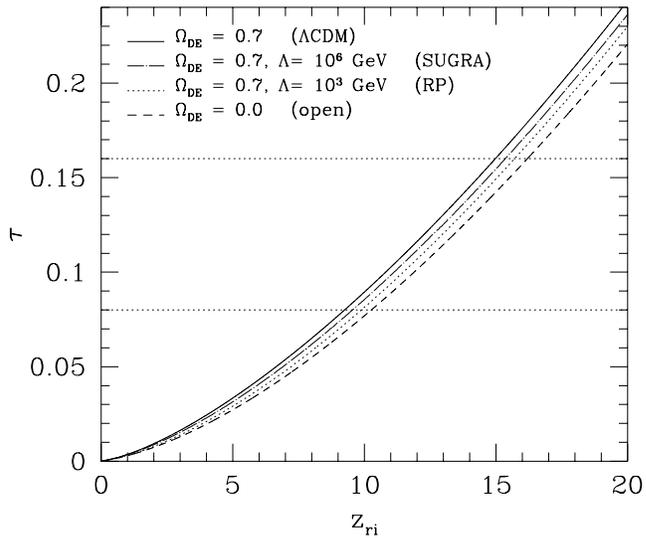}
\end{center}
\caption{Optical depth to CMB photons, in different models with $\Omega_m 
= 0.3$, $h=0.7$ and $\Omega_b h^2 = 0.022$, as a function
of the reionization redshift $z_{ri}$. The two horizontal dotted lines
show the reionization redshift consistent with $\tau=0.08$ or 0.16,
in the case of a single, fast and complete reionization event. }
\label{1fig}
\end{figure}

\begin{figure}
\begin{center}
\includegraphics*[width=9cm]{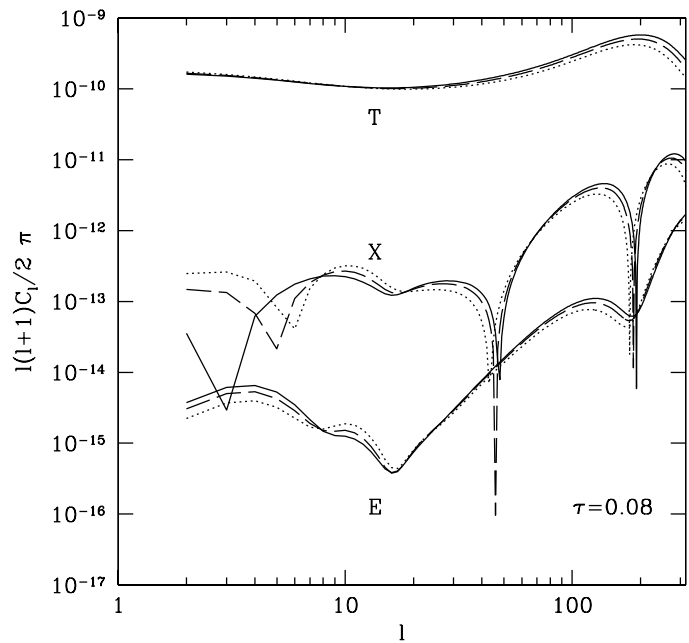}
\end{center}
\caption{Angular spectra of RP models with $\tau=0.08$. Continuous
(dashed, dotted) lines correspond to $\Lambda = 10^3~~(10^6,10^9)$GeV.
The absolute values of $C_l^{X}$ is plotted; changes of sign are indicated
by cusps; notice cusps at low $l$.}
\label{2fig}
\end{figure}

\begin{figure}
\begin{center}
\includegraphics*[width=9cm]{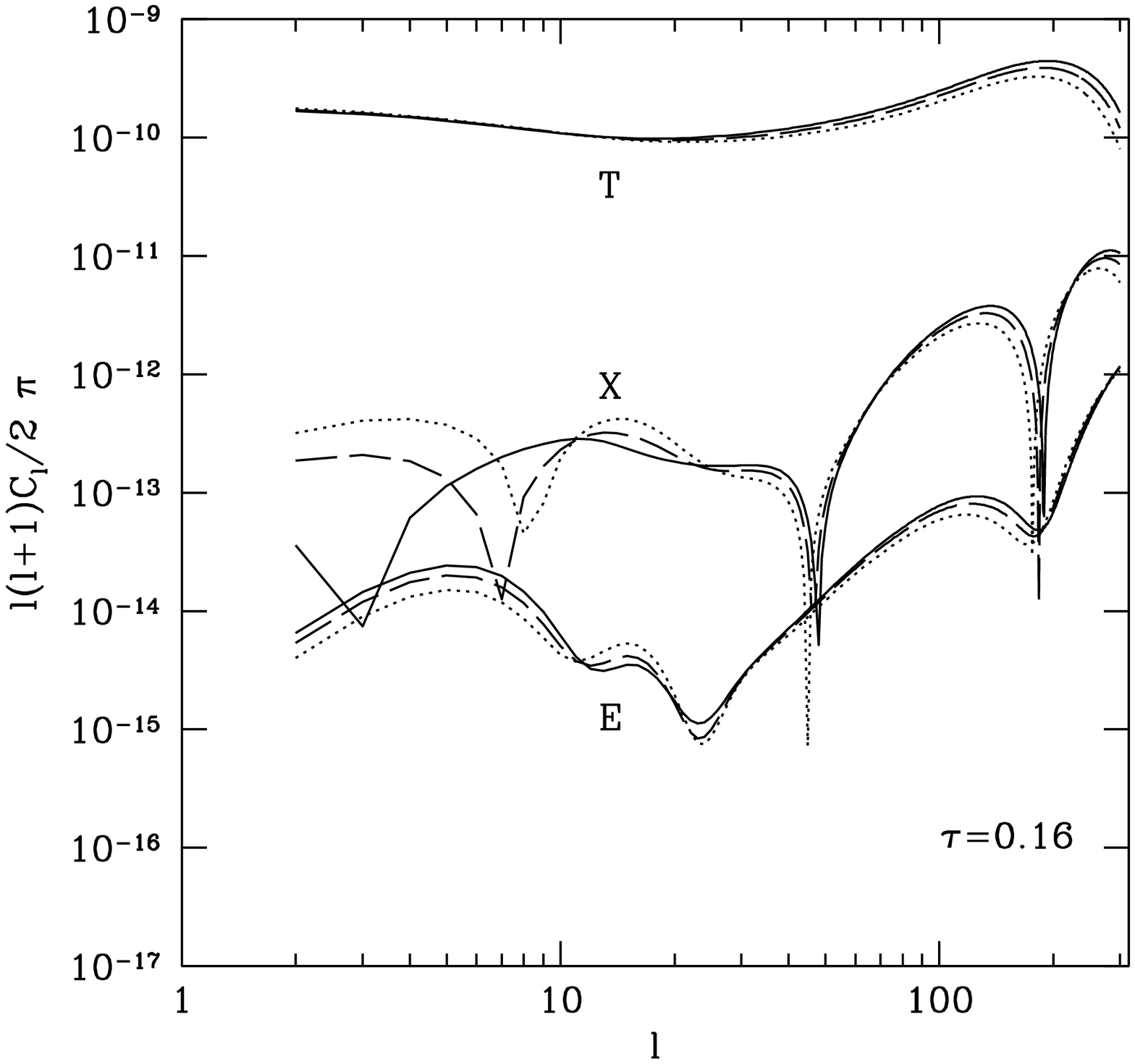}
\end{center}
\caption{As Fig.~2, but for $\tau = 0.16$.}
\label{3fig}
\end{figure}

\begin{figure}
\begin{center}
\includegraphics*[width=9cm]{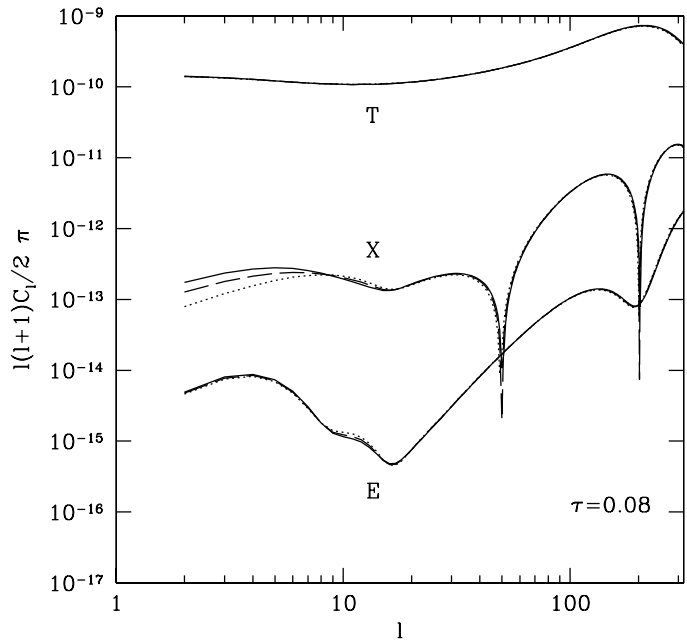}
\end{center}
\caption{As Fig.~2, but for SUGRA potentials. Also for 
SUGRA potentials, the main dependence on the energy scale
$\Lambda$ concerns the spectrum $C_l^{X}$ at quite low $l$.
This dependence is however much milder than for RP potentials.
}
\label{4fig}
\end{figure}

\begin{figure}
\begin{center}
\includegraphics*[width=9cm]{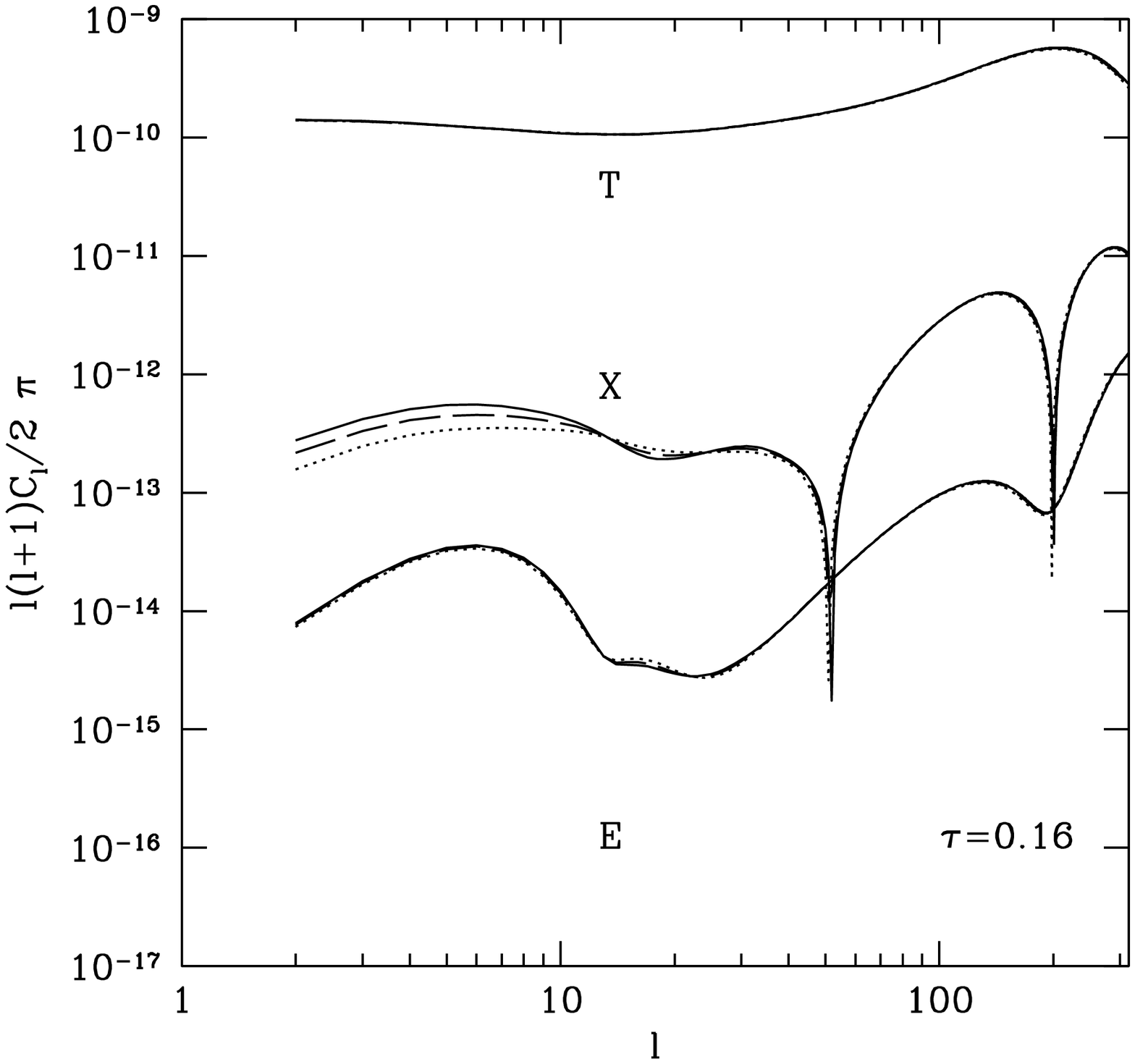}
\end{center}
\caption{As Fig.~4, but for $\tau = 0.16$.}
\label{5fig}
\end{figure}

\begin{figure}
\begin{center}
\includegraphics*[width=9cm]{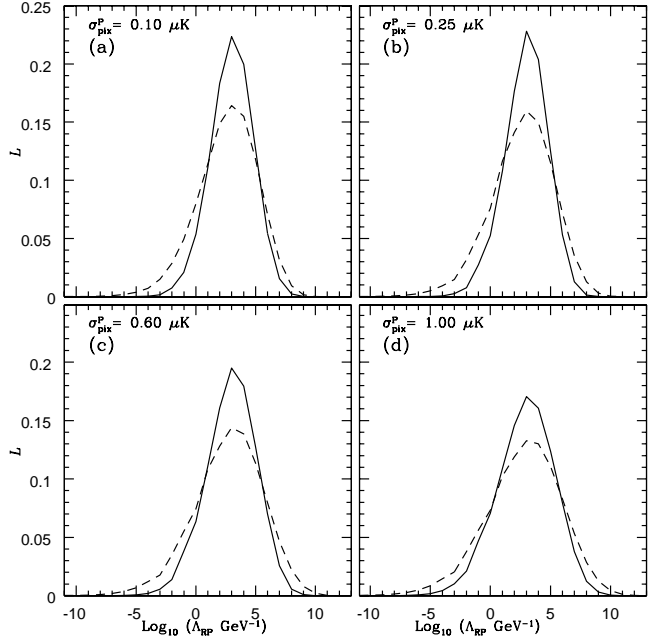}
\end{center}
\caption{The likelihood of various energy scales $\Lambda$ is determined
from artificial data, created for a model with given $\Lambda$ and for an
instrument measuring polarization with given $\sigma_{pix}^P$ (indicated
in the frames). In these plots $\Lambda= 10^3$GeV, continuous (dashed) 
lines refer to models with $\tau=0.16$ (0.08). The curves give the
likelihood obtained averaging among 1000 model realizations.
Even with $\sigma^P_{pix} \simeq 0.1\, \mu$K, there is a significant
likelihood for $\Lambda < 1\, $GeV. See Figs. \ref{13fig} to relate
such likelihood to the fraction of realizations for which the model
cannot be distinguished from $\Lambda$CDM.
}
\label{6fig}
\end{figure}

\begin{figure}
\begin{center}
\includegraphics*[width=9cm]{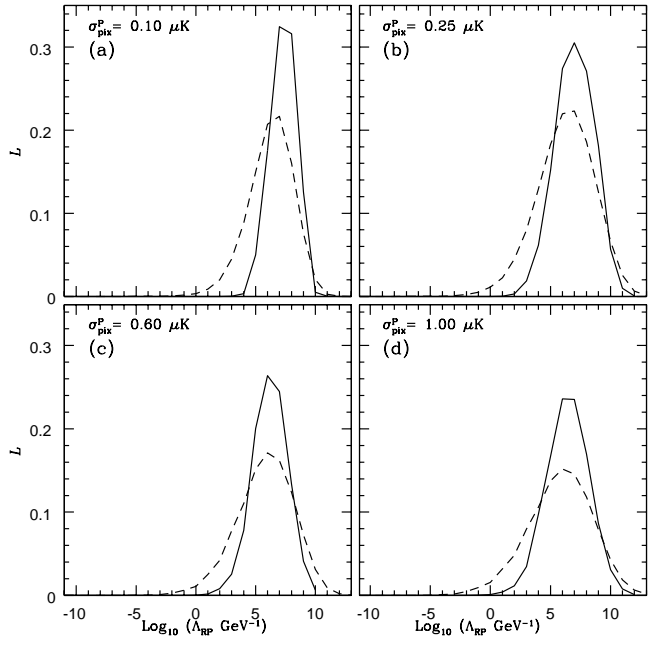}
\end{center}
\caption{The same as Fig.~6, but for $\Lambda = 10^6$GeV. Such model is only 
in marginal agreement with data on SNIa.}
\label{7fig}
\end{figure}


\begin{figure}
\begin{center}
\includegraphics*[width=9cm]{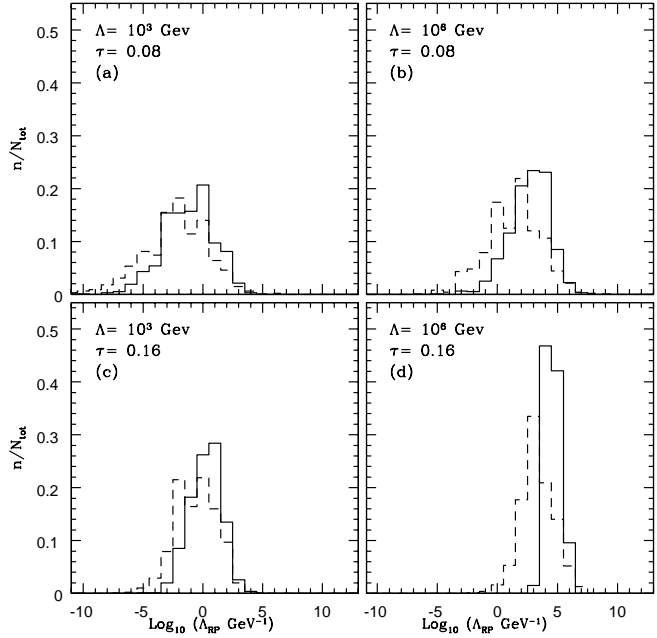}
\end{center}
\caption{In this (and the following) plot(s) we describe the effects 
of cosmic variance, showing how  1--$\sigma$ lower limits on $\Lambda$ 
depend on the model realization. The continuous (dashed) curve refer to 
$\sigma^P_{pix}=0.1~(1)\, \mu$K. In the frames, the values of the energy 
scale $\Lambda$ and of the opacity $\tau$ are indicated. Notice that, 
even in the worst case, there is a substantial fraction of realizations 
which allow to set a limit $\Lambda \magcir 1\, $GeV, so distinguishing
a RP model from $\Lambda$CDM. For $\Lambda=10^6$GeV and $\tau=0.16$,
the fraction of realizations indistinguishable from $\Lambda$CDM, 
with the reasonable noise level $\sigma_{pix}^P = 1\, \mu$K, is $\sim 2\, \%$.
}
\label{9fig}
\end{figure}

\begin{figure}
\begin{center}
\includegraphics*[width=9cm]{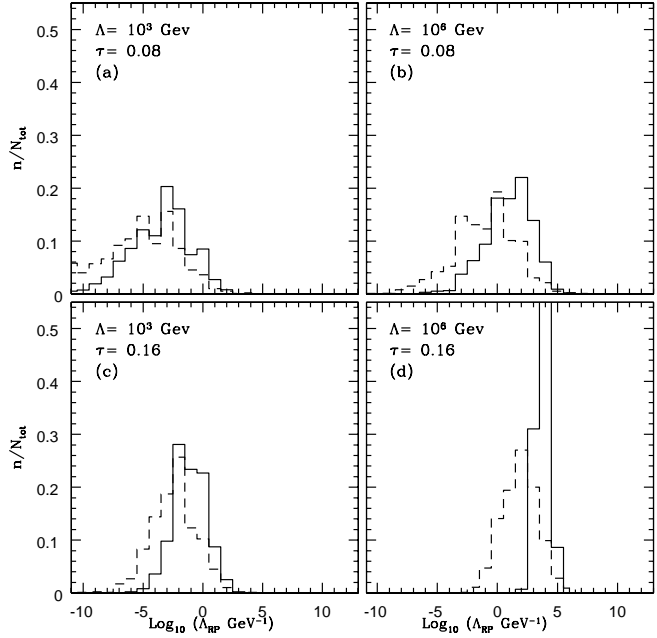}
\end{center}
\caption{As Fig.~8, at the 2--$\sigma$ confidence level.
For $\Lambda=10^{\, 6}$GeV and $\tau=0.16$, the fraction of realizations
indistinguishable from $\Lambda$CDM, with 
$\sigma_{pix}^P = 1\, \mu$K, is $\sim 16\, \%$.
}
\label{10fig}
\end{figure}


\begin{figure}
\begin{center}
\includegraphics*[width=9cm]{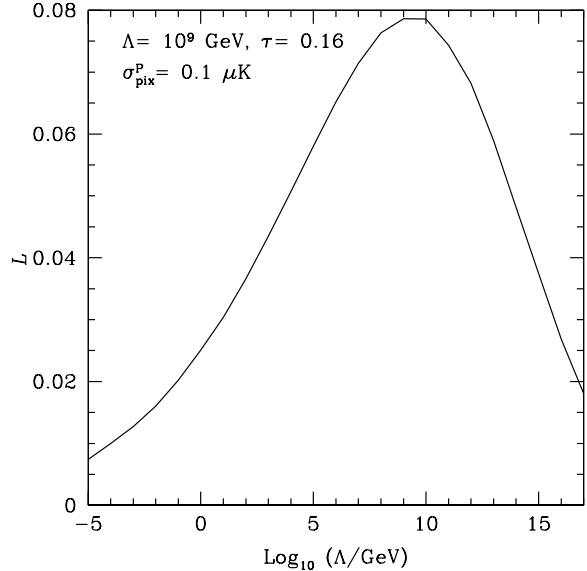}
\end{center}
\caption{This figure is analogous to Fig.~6 and concerns 
the SUGRA model for which $\Lambda$ is best determined by
artificial data. The pixel noise is 10 times smaller than the
most optimistic expected noise for SPOrt.
This plot is obtained by averaging over 1000 realizations.
}
\label{12fig}
\end{figure}

\begin{figure}
\begin{center}
\includegraphics*[width=9cm]{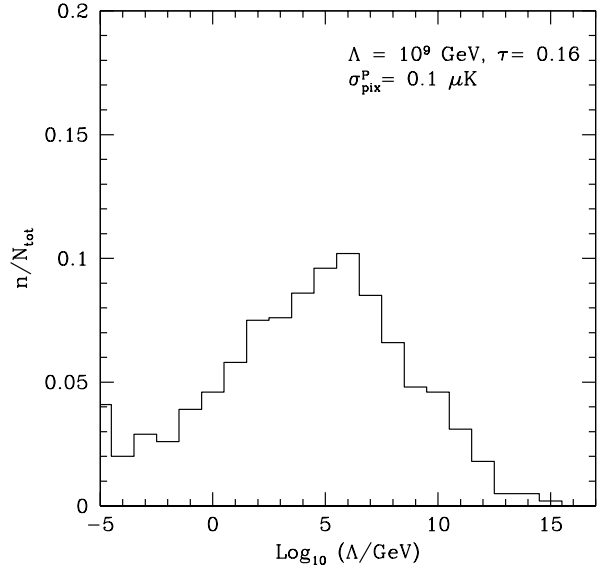}
\end{center}
\caption{1--$\sigma$ lower limit distribution for the SUGRA 
model  for which $\Lambda$ is best determined by
artificial data.
}
\label{13fig}
\end{figure}

\begin{figure}
\begin{center}
\includegraphics*[width=9cm]{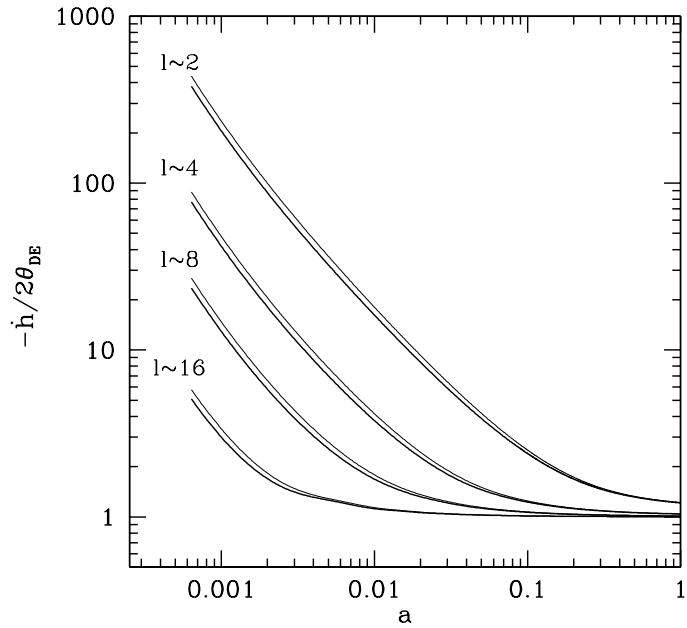}
\end{center}
\caption{The ratio $-\dot h/2\theta_{DE}$ is plotted with
a thick (thin) line for RP models with $\Lambda = 10^3 (10^6)$GeV
}
\label{14fig}
\end{figure}

\begin{figure}
\begin{center}
\includegraphics*[width=9cm]{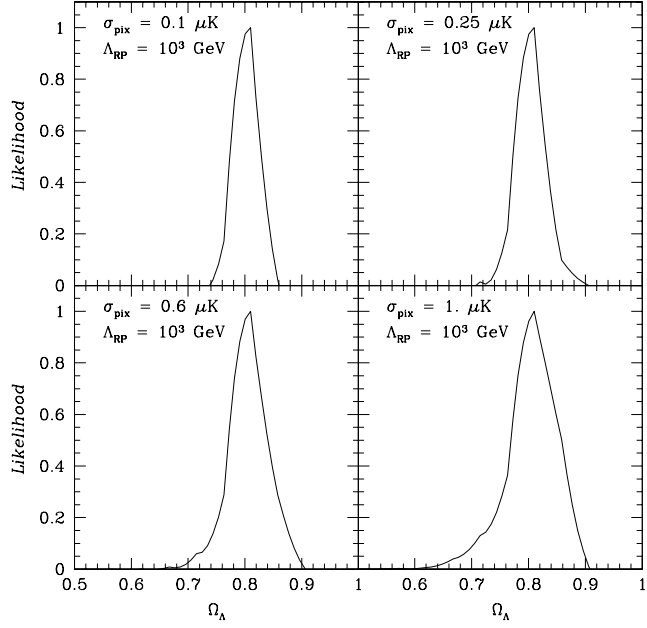}
\end{center}
\caption{Likelihood distribution on $\Omega_{DE} \equiv
  \Omega_\Lambda$, 
when  RP artificial data are fitted with $\Lambda$CDM models. RP
model parameters are shown in the frames. In all cases the true model
  has $\Omega_{DE} = 0.7$.
}
\label{ll3fig}
\end{figure}


\begin{figure}
\begin{center}
\includegraphics*[width=9cm]{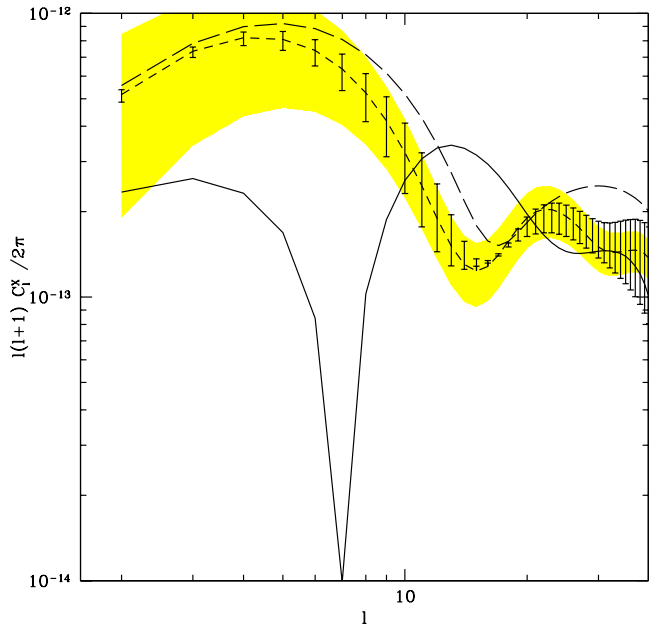}
\end{center}
\caption{$C^{X}_l$ spectrum of the $\Lambda$CDM model best fitting
RP model data with $\Lambda = 10^6$GeV. See text for more details.
}
\label{cxlfig}
\end{figure}

\begin{figure}
\begin{center}
\includegraphics*[width=9cm]{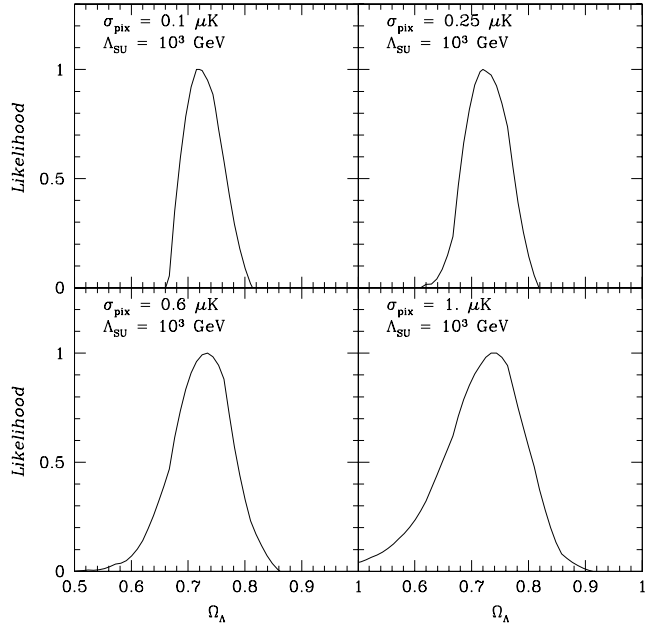}
\end{center}
\caption{Likelihood distribution on $\Omega_{DE}$, when SUGRA
artificial data are fitted with $\Lambda$CDM models. In this case,
the model parameters, shown in the frames, are well approached.
}
\label{llsufig}
\end{figure}

\begin{figure}
\begin{center}
\includegraphics*[width=9cm]{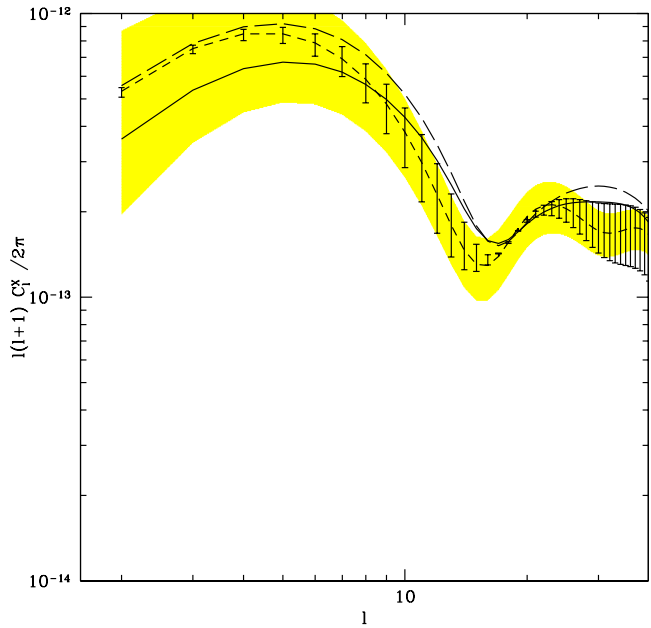}
\end{center}
\caption{$C^{X}_l$ spectrum of the $\Lambda$CDM model best fitting
SUGRA model data with $\Lambda = 10^3$GeV. See text for more details.
In this case, thanks to the lack of negative $TE$ correlation,
the fitting procedure has no serious misleading effect.
}
\label{cxlsufig}
\end{figure}

\end{document}